\documentclass[doublecol]{epl2} 
\usepackage{changepage}

\usepackage[utf8]{inputenc}

\usepackage{textcomp,marvosym}

\usepackage{fixltx2e}

\usepackage{amsmath,amssymb}
\usepackage{appendix}
\usepackage{cite}

\usepackage{nameref,hyperref}

\usepackage[right]{lineno}

\usepackage{microtype}
\DisableLigatures[f]{encoding = *, family = * }

\usepackage{rotating}

\title{\textbf{Slip of grip of a molecular motor on a crowded track: Modeling shift of reading frame of ribosome on RNA template}}
\shorttitle{Title} %Insert here a short version of the title if it exceeds 70 characters

\author{Bhavya Mishra\inst{1} \and Gunter M. Sch\"utz\inst{2} \and Debashish Chowdhury\inst{1}}
\institute{                    
  \inst{1} Department of Physics, Indian Institute of Technology,
Kanpur 208016, India.\\
  \inst{2} Institute of Complex Systems II, Forschungszentrum J\"ulich, 52425 J\"ulich, Germany.
}
\pacs{87.16.-b}{Subcellular structure and processes}
\pacs{05.40.-a}{Fluctuation phenomena, random processes, noise, and Brownian motion}
\pacs{05.70.Ln}{Nonequilibrium and irreversible thermodynamics}

\abstract{
We develop a stochastic model for the programmed frameshift of ribosomes synthesizing
a protein while moving along a mRNA template. Normally the reading frame of a ribosome 
decodes successive triplets of nucleotides on the mRNA in a step-by-step manner.
We focus on the programmed shift of the ribosomal reading frame, forward or backward, 
by only one nucleotide which results in a fusion protein; it occurs when a ribosome temporarily
loses its grip to its mRNA track. Special ``slippery'' sequences of nucleotides and also
downstream secondary structures of the mRNA strand are believed to play key roles in
programmed frameshift. Here we explore the role of an hitherto neglected parameter in
regulating -1 programmed frameshift. Specifically, we demonstrate that the frameshift
frequency can be strongly regulated also by the density of the ribosomes, all of which are
engaged in simultaneous translation of the same mRNA, at and around the slippery
sequence. Monte Carlo simulations support the analytical predictions obtained from a
mean-field analysis of the stochastic dynamics.\\
\noindent {\bf Key words}: ribosome traffic, master equation, programmed frame shift.
}

\begin{document}

\maketitle

\section{Introduction}
%%%%%%%%%%%%%%%%%%%%%%%%%%%%%%%%%%%%%%%%%%%%%%
A protein is a linear hetero-polymer made of a sequence of monomeric subunits, called
amino acids, each of which is linked to its immediate neighbor by a peptide bond. 
%For obvious reasons, a protein is also called a polypeptide. 
Nature normally uses 20 different 
types of amino acids to make proteins in living cells. The particular sequence of the types of amino 
acids in a protein is determined by the sequence of nucleotides, the monomeric subunits, 
of the corresponding template mRNA molecule. The actual synthesis of the protein, as 
directed by the mRNA template, is carried out by a molecular machine \cite{frank10}, 
called ribosome \cite{chowdhury13a,chowdhury13b,kolomeisky13} 
and the process is referred to as translation (of genetic message). 
Translation is broadly divided into three stages: Initiation, elongation and termination. 
Elongation of the growing protein by the ribosome takes place in a step-by-step manner,
the addition of each amino acid monomer to it is accompanied by a forward stepping of the  
ribosome on its mRNA template by one codon, each codon being a triplet of nucleotides. 
Thus, a ribosome is also a molecular motor \cite{chowdhury13a,kolomeisky13} 
that exploits the mRNA template as its track and moves forward along it, by three
nucleotides in each step, converting chemical energy into mechanical work.

At each position of the ribosome its ``reading frame'' decodes a triplet of nucleotides on
the mRNA template and then slides to the next triplet as the ribosome steps forward by
a codon. This reading frame is established in the initiation stage of translation and must
be maintained faithfully during the course of normal elongation of the protein. However,
in all kingdoms of life, on many template mRNA strands there are some special ``slippery''
sequences of nucleotides where a ribosome can lose its grip on its track, resulting in a shift
of its reading frame either backward or forward by one or more nucleotides. These processes
are referred to as ribosomal frameshift \cite{farabaugh97,atkins10}. The most commonly occurring, and extensively
studied, cases correspond to a shift of the reading frame backward or forward by a single
nucleotide on the mRNA track. These are referred to as -1 frameshift and +1 frameshift,
respectively, and will be the main focus of our study in this letter.

\begin{figure}[t] 
\begin{center}
\includegraphics[width=0.95\columnwidth]{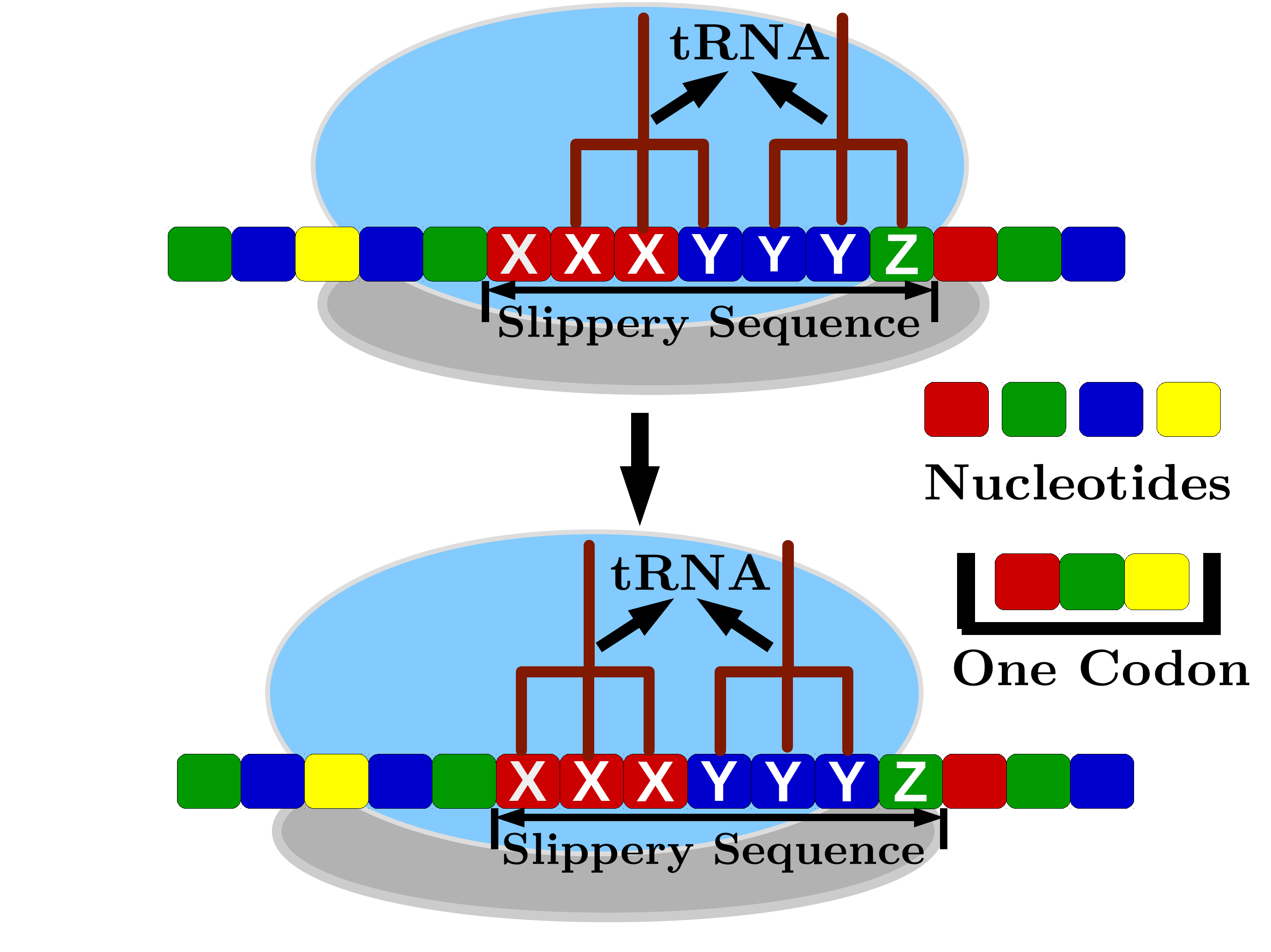} 
\end{center} 
\caption[scale=tiny]{(Color online) A schematic representation of -1 programmed ribosomal frameshift (PRF) from initial reading frame to -1 reading frame.} 
\label{fig-kinetics}
\end{figure}

The rate of frameshift at any arbitrary position on the mRNA track has been found
to be negligibly small. In contrast to such a random frameshift, a ``programmed'' frameshift at
a specific location on the mRNA track is known to occur with much higher {\it rate} and
have important biological functions. In this letter we consider, exclusively, programmed 
frameshift and ignore random frameshift altogether. After suffering a programmed frameshift, the ribosome
resumes its operation but decodes the template using the shifted reading frame (see Fig.~(\ref{fig-kinetics}))
thereby producing a ``fusion'' protein. The classic example of such a fusion product of -1 frame 
shift is the gag-pol fusion protein of the human immunodeficiency virus (HIV) \cite{farabaugh97,atkins10}.
Programmed frameshift, a mode of non-conventional translation \cite{firth12} happens to be one of
the several modes of genetic recoding \cite{atkins10}. A common feature of these recoding phenomena
is the context-dependent temporary alteration of the readout of mRNA. Understanding
the principles followed by nature for encoding and decoding genetic message would be
incomplete without understanding the causes and consequences of all such dynamic recoding.

%%%%%%%%%%%%%%%%%%%%%%%%%%%%%%%%%%%%%%%%%%%%%%
%%%%%%%%%%%%%%%%%%%%%%%%%%%%%%%%%%%%%%%%%%%%%%%

Programmed -1 frameshift requires two key ingredients: (a) A slippery sequence (usually
about seven nucleotide long) on the mRNA, and (b) a downstream secondary structure
(usually a pseudoknot \cite{brierley07,giedroc09}) of the mRNA 
\cite{kim14}. 
Normally the pseudoknot on the mRNA template is located about 6 nucleotides downstream from 
the slippery sequence. 
In order to enter the segment of mRNA template that forms the pseudoknot, a ribosome has 
to unwind the secondary structure. Thus, undoubtedly, the pseudoknot acts as a roadblock 
against the forward movement of the ribosome that suffers a long pause on the slippery site. 
However, such long pauses are necessary, but may not be sufficient, for inducing -1 frameshift 
of the ribosomes because not all types of road blocks on mRNA can induce frameshift \cite{brierley08}. 
Moreover, not every ribosome suffers frameshift at a given slippery sequence. 
Even more intriguing is the fact that the same pseudoknot, that puts such an insurmountable 
hurdle on the path of a ribosome in the 0 reading frame, allows it to pass through more easily 
after the shift to the -1 reading frame.

Thus it is evident that the composition and
length of the slippery sequence, its distance from the downstream pseudoknot, the energetic
stability of the pseuodoknot and the kinetics of its unfolding and refolding etc. collectively
encode the program that determines not only the spatial location and timing of the frame
shift but also its frequency or efficiency. A combination of all these stimulators and signals
are believed to alter the normal free energy landscape of the ribosome, thereby affecting the
stability of various intermediate states as well as the kinetics of transitions among them that favor frameshift over the other alternative pathways \cite{farabaugh97,atkins10}.
Several competing models have been developed to account for the mechanisms of stimulation, regulation and control of frameshift; the models differ in (a) their hypothesis as to
the sub-step of the mechano-chemical cycle in which the frameshift is assumed to occur, and
(b) the assumed structural, energetic and kinetic cause of the slippage 
\cite{tinoco13}.
Instead of committing at present to any 
specific structural model for frameshift, we capture the effects of the slippery sequence and the 
downstream pseudoknot by physically motivated generic alterations of the kinetic rates in a 
reduced minimal model of the elongation kinetics.

The main aim of this letter is to demonstrate how an hitherto neglected control parameter, 
namely the average inter-ribosome separation, or equivalently the mean number density
of the ribosomes, on the mRNA track can up- or down-regulate the efficiency of ribosomal
frameshift {\it in vivo}. In the past indications for the interplay of the inter-ribosome separation
and the kinetics of unwinding of the pseudoknot have emerged from experimental studies
of frameshift {\it in vivo} \cite{lopinski00} as well as in experiments with synthetic mRNA 
secondary structures {\it in vitro} \cite{tholstrup12}.

Because of the superficial similarities between the simultaneous movement of multiple
ribosomes along a single mRNA track and vehicular traffic along a single-lane highway, 
the former is often referred to as ribosome traffic. Various aspects of this traffic-like 
phenomenon have been modelled by totally asymmetric simple exclusion process 
(TASEP) \cite{schutz00}
and its biologically motivated extensions. The TASEP, an abstract model of self-driven interacting 
particles, is one of the simplest models of driven non-equilibrium systems in statistical physics.
Since the pseudoknot acts as a bottleneck against the flow of ribosome traffic,
the spatio-temporal organization of the ribosomes exhibit some of the key characteristics
of TASEP with quenched defects; these include, as we show here, phase segregation of the
ribosomes. 

In contrast to the earlier TASEP-based models of translation 
\cite{css,scn,chowdhury05,sharma11a,sharma11b,chou11,zia11,haar12,rolland15,chou04,lakatos03,mitarai13,turci13} 
the model introduced here treats 
individual nucleotides, rather than triplets of nucleotides (codons), as the basic unit of the 
mRNA track. Moreover, unlike the earlier TASEP-based models of translation, we do not 
focus on the phase diagram of the model. Instead,the most important result of our investigation 
is the up- and down-regulation of ribosomal frameshift caused by the long queue that stretches 
upstream from the pseudoknot to the slippery sequence and beyond. We also suggest new 
applications of existing experimental techniques to test some of the new predictions made in 
this paper on the basis of our theoretical calculations.

%%%%%%%%%%%%%%%%%%%%%%%%%%%%%%%%%%%%%%%%%%%%%
\begin{figure}[t] 
\begin{center}
\includegraphics[width=0.95\columnwidth]{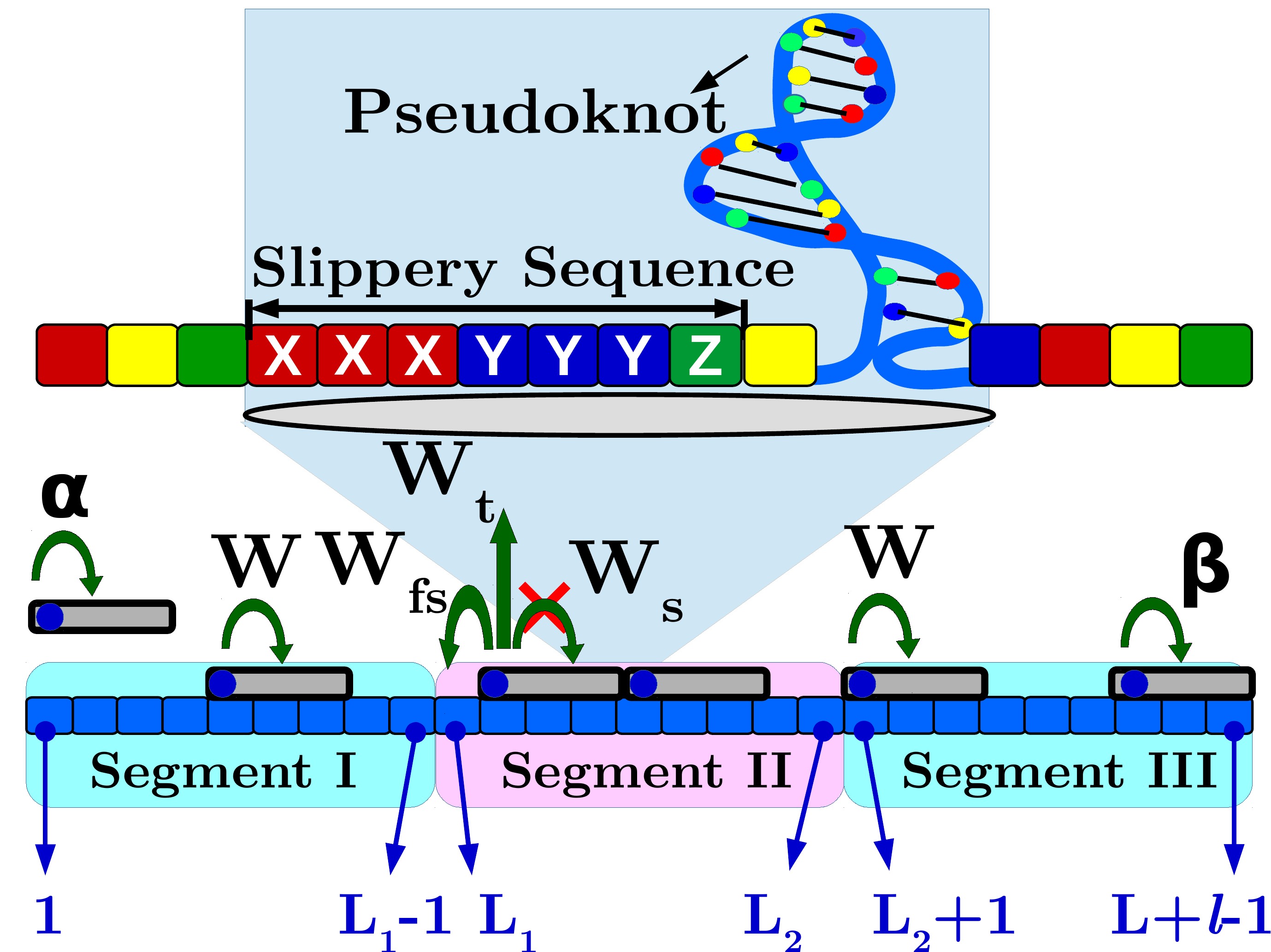}  
\end{center} 
\caption[scale=tiny]{(Color online) A schematic depiction of the model where a secondary structure of the mRNA (a cartoon of a pseudoknot) is shown. The mapping of the tortuous contour of the mRNA template onto a linear (one-dimensionl) chain is shown explicitly. Each site of the chain represents a single nucleotide. The first seven sites of the segment II form a ``slippery'' sequence while the next 2-3 nucleotides would correspond to the spacer region whereas the remaining part of the segment II remains folded in the pseudoknot. From the specific site $L_{1}+1$ programmed -1 frameshift can take place. The segments I and III are the segments of the mRNA that are located before and after, respectively, of the pseudoknot.  } 
\label{modelA}
\end{figure}
%%%%%%%%%%%%%%%%%%%%%%%%%%%%%%%%%%%%%%%%%%%%%

%%%%%%%%%%%%%%%%%%%%%%%%%%%%%%%%%%%%%%%%%%%%%%%%
\begin{figure}[t] 
\begin{center}
\includegraphics[width=0.95\columnwidth]{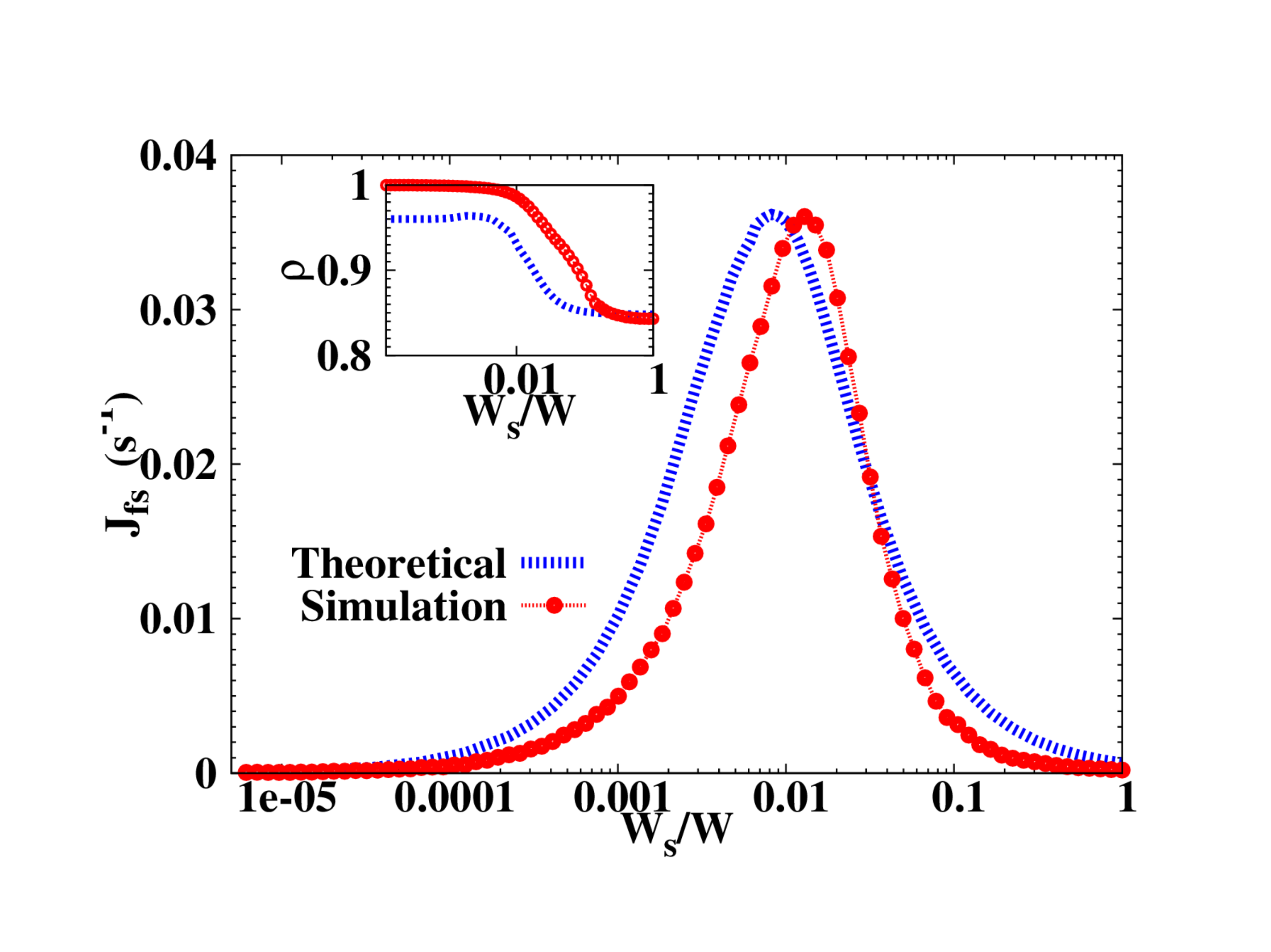}  
\end{center} 
\caption[scale=tiny]{(Color online) Frameshift flux $ J_{fs} $ is plotted against the parameter $W_{s}/W$. In the inset, the probability of coverage of the site $ i=L_{1}$,  which is targetted for successful -1 frameshift, is plotted also against $W_{s}/W$. The theoretical predictions derived under MFA are drawn by the dashed curve (blue). The numerical data obtained from MC simulations have been plotted with dots; the curve connecting these dots (red) serves merely as a guide to the eye. The initiation rate $ \alpha=100s^{-1} $, termination rate $ \beta=10s^{-1} $ and $W = 83$ s$^{-1}$. The system size is $N= L+ \ell-1=1000+\ell-1 $, with the rod size $ \ell=18 $, $L_{1}=399, L_{2}=450$.}
\label{eff1}
\end{figure}
%%%%%%%%%%%%%%%%%%%%%%%%%%%%%%%%%%%%%%%%%%%%%%%%%

%%%%%%%%%%%%%%%%%%%%%%%%%%%%%%%%%%%%%%%%%%%%%%%%%%
\begin{figure}[t] 
\begin{center}
\includegraphics[width=0.85\columnwidth]{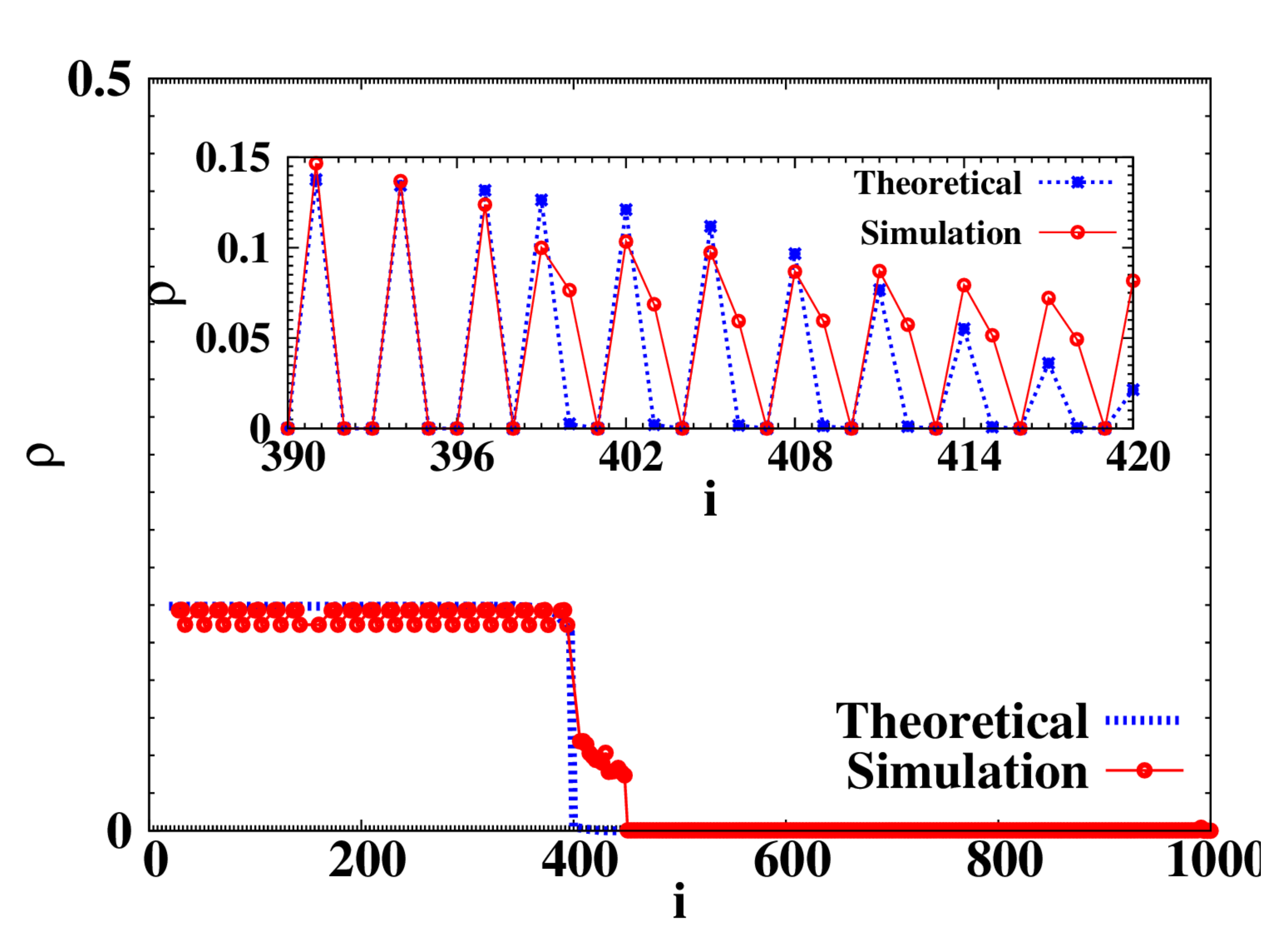}  
\end{center} 
\caption[scale=tiny]{(Color online) Occupational density profile in steps of 3 nucleotides. The dashed line (blue) is the theoretical prediction under MFA and the solid line (red) connecting the discrete symbols $\odot$ is obtained from MC simulation with initiation rate $ \alpha=100s^{-1} $, termination rate $ \beta=10s^{-1} $,  $W=83$s$^{-1}$,  $W_{fs}=13.3s^{-1}$ and $W_{s}=0.00074s^{-1}$. System size {$ N=L+\ell-1=1000+\ell-1 $} with the rod size {$ \ell=18 $}; the special site is located at $i=400$. The inset shows the occupational density profile of a small region from $ i=390 $ to $ i=420 $ in steps of 1 nucleotide; the dashed line (blue) with $ \ast$ and the solid line (red) connecting the discrete symbols $\odot$ have been obtained from MF theory and MC simulation, respectively.}
\label{eff3}
\end{figure}
%%%%%%%%%%%%%%%%%%%%%%%%%%%%%%%%%%%%%%%%%%%%%%%%%%

%%%%%%%%%%%%%%%%%%%%%%%%%%%%%%%%%%%%%%%%%%%%%%%%%%
\begin{figure}[t] 
\begin{center}
\includegraphics[width=0.85\columnwidth]{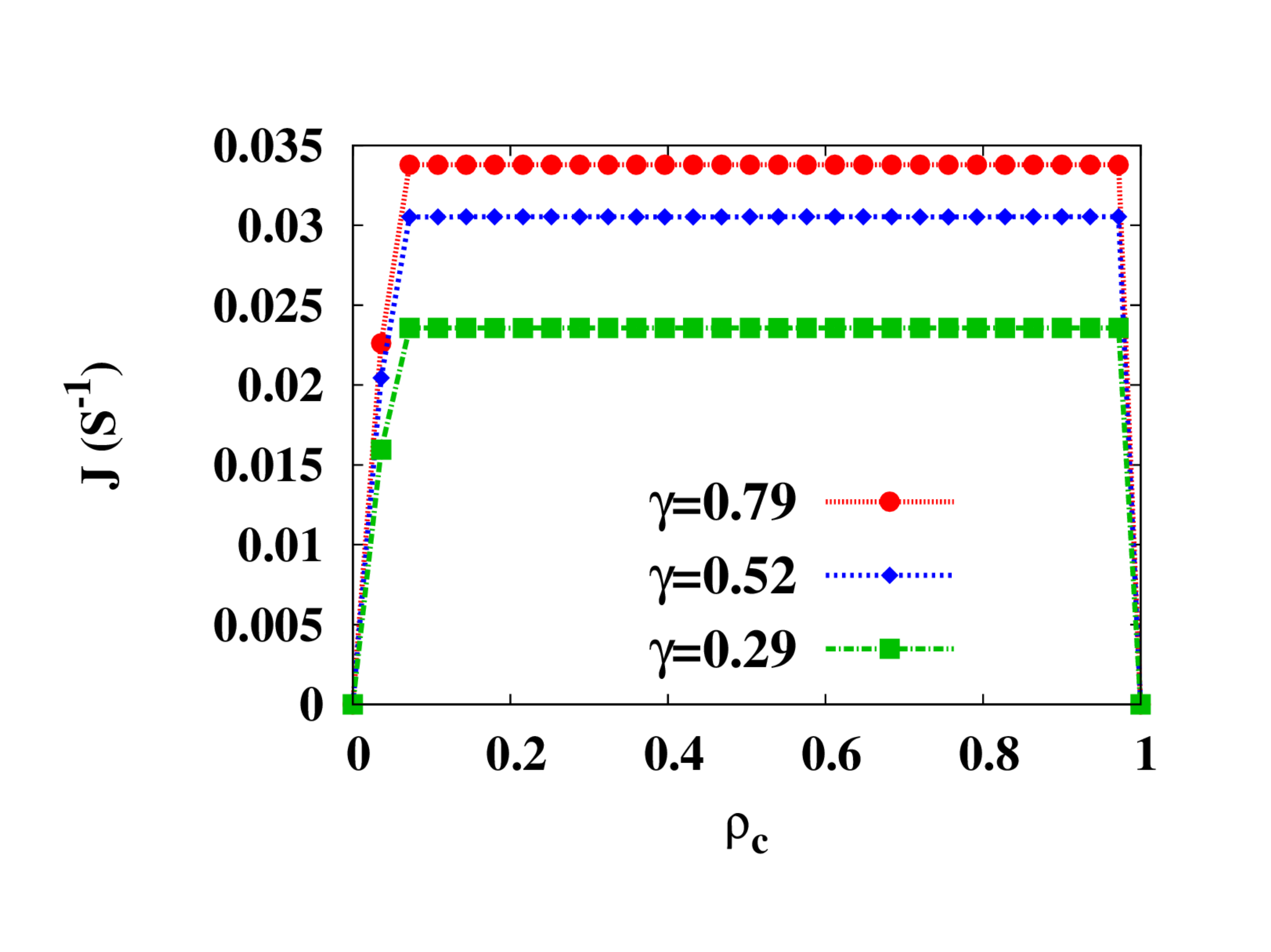}  
\end{center} 
\caption[scale=tiny]{(Color online) Net forward flux (J) plotted against the coverage density $\rho_{c}$ for three different values of the jump rate ratio $ \gamma = W_{s}/W$. System size is $ N=1000+\ell-1 $, with the rod size $ \ell=18 $, and $W=83$ s$^{-1}$.} 
\label{flux1}
\end{figure}
%%%%%%%%%%%%%%%%%%%%%%%%%%%%%%%%%%%%%%%%%%%%%%%%%%

%%%%%%%%%%%%%%%%%%%%%%%%%%%%%%%%%%%%%%%%%%%%%%
\section{Model}
%%%%%%%%%%%%%%%%%%%%%%%%%%%%%%%%%%%%%%%%%%%%%%
In our model mRNA is treated as a linear chain (also called a lattice) of 
sites,
each of which represents a single nucleotide. Each nucleotide is identified with an integer 
index $j$. The total length of the chain is {$L+{\ell}-1$},  in the units of nucleotide length, 
although only the segment starting from $j=1$ to $j=L$ gets translated by the ribosomes. 
A specific site $S$ on this chain denotes the {\it second} nucleotide of the slippery sequence 
\cite{atkins10}. 
A ribosome is modelled as a rigid rod of length {$\ell$} also in the units of nucleotide length, 
i.e., each ribosome can cover simultaneously  {$\ell $} successive nucleotides on its mRNA 
template. According to the convention followed consistently throughout this paper, the 
instantaneous position of a ribosome on the mRNA template is denoted by the integer index 
that labels the leftmost site {\it covered} by the ribosome at that instant of time. We make 
a clear distinction between the terms {\it occupied} and {\it covered}: A site $j$ is {\it occupied} 
by a ribosome, i.e., its instantaneous position is $j$, if the ribosome is decoding the triplet 
of nucleotides $j, j+1, j+2$, while all the sites {$j,j+1,...j+{\ell}-1$} remain {\it covered} by it 
simultaneously at that instant.

In this generic model developed here for ribosomal frameshift the lattice is divided into three 
segments. The segment $I$ is ranging from site $1$ to site $L_{1}-1$ ($1 \leq j \leq L_{1}-1$), 
segment $II$ from site $L_{1}$ to site $L_{2}$ ($L_{1} \leq j \leq L_{2}$) and segment $III$ 
from site $L_{2}+1$ to site {$L+{\ell}-1$} ({$L_{2}+1 \leq j \leq L+{\ell}-1$}) (see 
Fig.~\ref{modelA}). The segment II 
represents the stretch of the mRNA template that is folded in the form of the pseudoknot.
All the numerical data presented here have been obtained with the choice $ L_{1}=399$ and 
$ L_{2}=450 $. Inside segment $II$ seven nucleotides, from site $ j=L_{1} $ to $ j=L_{1}+6 $ 
represent a slippery sequence whose second site (i.e., $j=L_{1}+1$) is the special site from where 
the ribosomal frameshift is assumed to take place. The next 2-3 nucleotides would correspond 
to the spacer region between the slippery sequence and the pseudoknot while the remaining 
nucleotides of segment II form the pseudoknot. The choice of the numerical value $ 41 $ nucleotide for the length of the mRNA in the pseudoknot is only a typical one that lies between estimated lengths of the  smallest and largest pseudoknots \cite{giedroc09}. 

Translation initiation is captured in this model as follows:  If the first {$\ell$} sites of the lattice 
are empty, a ribosome can occupy the position $j=1$ (and, thus, cover the sites {$1,2,...,{\ell}$}).
This event occurs with a rate (i.e., probability per time unit) $\alpha$. Similarly, termination 
of translation is described as the detachment of the ribosome from the lattice when its position 
is $L$, i.e., it {\it covers} the last {${\ell}$} sites of the lattice that are labelled by {$L, L+1,...,L+{\ell}-1$}; 
the rate of this event is $ \beta $. 

During the elongation stage a ribosome moves forward by three nucleotides upon successful 
completion of each elongation cycle. However, at the site $S$, which represents the {\it second} 
nucleotide of the slippery sequence, a ribosome can slip backward on its track by one single 
nucleotide; the rate of this -1 frameshift event, that we define below, is normally much 
less than unity. So far as the termination of translation by a frameshifted ribosome is 
concerned, we assume, for the sake of simplicity, that the ribosome detaches from the lattice 
when its position is $L-1$, i.e., it covers the sites {$L-1, L,...,L+{\ell}-2$}. Thus, in this model the 
{\it full length} of a protein synthesized using the non-shifted frame and that of a fusion protein 
would consist of the same number of amino acids.

In this scenario the overall rate of translation of each codon in the original (unshifted) frame is captured by an ``effective rate'' $W$  of ``hopping'' of a ribosome by three 
steps in the forward direction, i.e., from position $j$ to $j+3$. In principle, the effective rate $W$ 
can be expressed in terms of the actual rates of the individual transitions among the five distinct 
mechano-chemical states in the elongation stage (for details please see the supplementary information 
given in \cite{mishra16a}).

Next we assume that  the effective hopping rate $W$ in the pseudoknot segment $II$
gets reduced exponentially to the value $W_{s}$, i.e., 
%%%%%%%%%%%%%%%%%%%%%%%%%%%%%%%%%%%%%%%%%%%%%%
\begin{center}
\begin{equation}
W_{s}/W= \gamma= \exp(-b~\Delta \tilde{G})
\label{ratew}
\end{equation}
\end{center}
%%%%%%%%%%%%%%%%%%%%%%%%%%%%%%%%%%%%%%%%%%%%%%
where $ -\Delta \tilde{G} $ is a dimensionless parameter. This choice is motivated by the 
plausible identification $ \Delta \tilde{G} = (\Delta G)/(k_{B} T ) $ where $ \Delta G  $ is the free 
energy barrier against forward movement of the ribosome by a single nucleotide within the 
pseudoknot region. Since a typical pseudoknot is not a mere hairpin, the effects of its structural 
complexity on the effective barrier is captured by the parameter $b$ in the exponential. For the 
numerical data plotted graphically in this paper $b=3$ was chosen. As  
$\Delta \tilde{G} \to 0$ the choice of the form of $W_{s}$ in (\ref{ratew}) implies $W_{s} \to W$; 
this is consistent with the fact that in this limit the difference between the segment II and the 
other two segments disappears. 
In the opposite limit $\Delta \tilde{G} \to \infty$, $W_{s} \to 0$ as would be expected on physical 
grounds that it is practically impossible to unzip an extremely stiff psudoknot.
The rate of frameshift $ W_{fs} $ depends upon two factors: (a) Strength of the pseudoknot,  
expressed by $ \Delta \tilde{G} $, and (b) the frequency $W_{fs0}$ which is related to the 
breaking the bonds between tRNA and codons. 

Motivated by the above physical considerations we make our next postulate: We assume 
that the rate of -1 frameshift in the slippery region is given by 
%%%%%%%%%%%%%%%%%%%%%%%%%%%%%%%%%%%%%%%%%%%%%
\begin{center}
\begin{equation}
W_{fs}=W_{fs0}\exp(a~\Delta \tilde{G})
\label{ratewfs}
\end{equation}
\end{center}
%%%%%%%%%%%%%%%%%%%%%%%%%%%%%%%%%%%%%%%%%%%%%
where $a$ is the parameter that indirectly captures the complexity of pseudoknot structure. 
Note that in the limit $\Delta \tilde{G} \to 0$, $W_{fs} \to W_{fs0}$ which is a 
non-vanishing (but, presumably small) rate of frameshift caused by the slippery sequence in 
the absence of a downstream psudoknot. Moreover, the form (\ref{ratewfs}) is based on the 
assumption of a sharp increase of $W_{fs}$ with increasing $\Delta \tilde{G}$.
For the numerical data plotted graphically in this paper $a=3$ was chosen.\cite{brierley07}.

Similarly, the detachment  (premature termination) rate $ W_{t} $ of a rod from the 
special site is assumed to be
%%%%%%%%%%%%%%%%%%%%%%%%%%%%%%%%%%%%%%%%%%%%%%%%%%%%%%%%%%%%%%%%%
\begin{equation}
W_{t}=W \exp(- c~ \Delta \tilde{G} )
\label{ratewt}
\end{equation}
%%%%%%%%%%%%%%%%%%%%%%%%%%%%%%%%%%%%%%%%%%%%%%%
Rods can detach prematurely from the special site only if the nearest neighbour site from the 
rightmost part of the rod in the forward direction is already covered by another rod. 
In an alternative version of this model, defined in the supplementary material, the rods can 
detach prematurely if the nearest neighbor of the left most part of the rod in backward direction, 
i.e., the site targetted for -1 frameshift, is already covered by another rod. 
The choice of the form (\ref{ratewt}) is motivated by our postulate that (a) in the absence of the 
pseudoknot (i.e., in the limit $\Delta \tilde{G} \to 0$) both the forward hopping and premature 
detachment at the slippery site are equally probable, and that (b) for very stiff pseudoknots 
(i.e.,  $\Delta \tilde{G} \to \infty$) practically a ribosome stalls (no forward movement because 
$W_{s} \to 0$) and, therefore, no possibility of premature detachment \cite{chen14}. Although, in principle, 
the two parameters $a$ and $c$ in (\ref{ratew}) and (\ref{ratewt}) are not necessarily equal, 
we use $a=c$ just for the sake of simplicity.

%%%%%%%%%%%%%%%%%%%%%%%%%%%%%%%%%%%%%%%%%%%%% 
%%%%%%%%%%%%%%%%%%%%%%%%%%%%%%%%%%%%%%%%%%%%%
%%%%%%%%%%%%%%%%%%%%%%%%%%%%%%%%%%%%%%%%%%%%%
%\section{Model}

%%%%%%%%%%%%%%%%%%%%%%%%%%%%%%%%%%%%%%%%%%%%%%

The kinetics is implemented by the following rules:\\
(a) A new rod can attach at site $1$, if and only if all initial {$\ell$} sites are empty. The rate of attachment at site $1$ is $\alpha$.\\
(b) If there is a rod at site $i=L$ or $i=L-1 $, then it can detach from the lattice, and the rate of detachment is $\beta$.\\
(c) Inside segment $I$ and $III$ rods can jump in forward direction by $ +3 $ nucleotides only if the target site is empty and the rate of forward jump is $W$.\\
(d) Inside segment $II$ there is one special site $i=L_{1}+1$. Except at this special site a rod can jump forward, by step size $+3$, if the target site is empty and the rate of forward jump in segment $II$ is $W_s$.\\
(e) From the special site $i=L_{1}+1$ the following movements are possible:\\
(e1) A Rod can jump forward with rate $W_s$, with step size $+3$, if the target site is empty.\\
(e2) A Rod can slip back with step size $-1$ with rate $W_{fs}$, if the target site is empty.\\
(e3) A Rod can detach, with rate $W_{t}$,  from the lattice if the site {$L_{1}+1+{\ell}$} (i.e., the site immediately in front of its forward edge) is occupied by another rod.\\
 
%%%%%%%%%%%%%%%%%%%%%%%%%%%%%%%%%%%%%%%%%%%%%%
%%%%%%%%%%%%%%%%%%%%%%%%%%%%%%%%%%%%%%%%%%%%%%%
In the analytical treatment of this model
$ p(j|\underline{i}) $ is the conditional probability of finding the site $ j $ empty, given that the site $ i $ is already occupied, where $j$ is the target site when a rod tends to move by $\pm 1$ nucleotides.
Similarly, $ q(k|\underline{i}) $ is the conditional probability of finding the site $ k $ empty, given that the site $ i $ is already occupied, where $ k $ denotes the target site when the rod tends to move by $ 3 $ nucleotides.
%%%%%%%%%%%%%%%%%%%%%%%%%%%%%%%%%%%%%%%%%%%%%
$ P(i,t) $ is the occupational probability and it is defined as the probability of finding the left edge of the rod at site $ i $ at time $ t $.

%%%%%%%%%%%%%%%%%%%%%%%%%%%%%%%%%%%%%%%%%%%%%%

In this work we are interested in the frameshift flux ($J_{fs}$) from the special point which is defined as the total number of ribosomes that undergo frameshift per unit time from this site. Thus, $ J_{fs} $ is given by
%%%%%%%%%%%%%%%%%%%%%%%%%%%%%%%%%%%%%%%%%%%%%%%%
\begin{equation}
J_{fs}=W_{fs}p(L_{1}|\underline{L_{1}+1})P(L_{1}+1).
\label{frameshiftfluxA}
\end{equation} 

%%%%%%%%%%%%%%%%%%%%%%%%%%%%%%%%%%%%%%%%%%%%%%%%

We can get the value of $ P(i,t) $ at each individual site $ i $ and time $ t $ by solving master equations under mean field approximation (MFA). 
%%%%%%%%%%%%%%%%%%%%%%%%%%%%%%%%%%%%%%%%%%%
At site $ \textit{i}=1 $ one has 
\begin{equation}
{
\begin{split}
\dfrac{dP(i,t)}{dt} &= \alpha [1-\sum_{a=1}^{\ell}P(a,t)]\\
& -Wq(i+ \ell+2|\underline{i})P(i,t),
\end{split}
}
\label{mfeq1}
\end{equation} 
%%%%%%%%%%%%%%%%%%%%%%%%%%%%%%%%%%%%%%%%%%%%%%%%%
since a new rod can attach only when all initial {$ \ell $} sites are empty. 
Therefore  a summation up to 
{$ \ell $} is taken in the gain part of Eq.~\ref{mfeq1}. 

%%%%%%%%%%%%%%%%%%%%%%%%%%%%%%%%%%%%%%%%%%%%%%%%%
At the special site $ \textit{i}=L_{1}+1 $ one gets
\begin{equation}
{
\begin{split}
\dfrac{dP(i,t)}{dt} & = W_{s}q(i+ \ell-1 |\underline{i-3})P(i-3,t) \\
& -[W_{s}q(i+ \ell+2|\underline{i})+W_{fs}p(i-1|\underline{i})]P(i,t) \\
& -W_{t}P(i+ \ell,t)P(i,t),
\end{split}
}
\label{mfeq2}
\end{equation}
%%%%%%%%%%%%%%%%%%%%%%%%%%%%%%%%%%%%%%%%%%%%%%%%%%%%%%%%%
and at site $ \textit{i}=L_{1} $
\begin{equation}
{
\begin{split}
\dfrac{dP(i,t)}{dt} &= W_{s}q(i+ \ell-1 |\underline{i-3})P(i-3,t) \\
& +W_{fs}p(i|\underline{i+1})P(i+1,t) \\
& -W_{s}q(i+ \ell+2|\underline{i})P(i,t).
\end{split}
}
\label{mfeq3}
\end{equation}
%%%%%%%%%%%%%%%%%%%%%%%%%%%%%%%%%%%%%%%%%%%%%%%%%%%%%%%%%
At the site of termination of translation 
\begin{equation}
\dfrac{dP(i,t)}{dt}= WP(i-3,t)-P(i,t) \beta,
\label{mfeq4}
\end{equation}
where $i=L$ for non-frameshifted ribosomes where $i=L -1$ in case of frameshifted ribosome. Note that when the rod is located at $i=L-4$ or at $i=L-3$ it does not face any exclusion at its 
target site $L-1$ or $L$, respectively; consequently, there is no factor of $q$ in the gain term on the 
right hand side of  (\ref{mfeq4}).

%%%%%%%%%%%%%%%%%%%%%%%%%%%%%%%%%%%%%%%%%%%%%%%%%%%%%%%%%%%%%%
%%%%%%%%%%%%%%%%%%%%%%%%%%%%%%%%%%%%%%%%%%%%%%%%%%%%%%%%%%%%%%%
For the same reason, for all the sites {$ i \geq L-\ell+1 $}, exclusion effect appears neither in the gain terms nor in the loss terms in the master equations 
\begin{equation}
\dfrac{dP(i,t)}{dt}= WP(i-3,t)-WP(i,t) .
\label{mfeq4a}
\end{equation}
%%%%%%%%%%%%%%%%%%%%%%%%%%%%%%%%%%%%%%%%%%%%%%%%%%%%%%%%%%%%%%\
%%%%%%%%%%%%%%%%%%%%%%%%%%%%%%%%%%%%%%%%%%%%%%%%%%%%%%%%%%%%%%%
At all other sites in segment $I$ and $III$\\
\begin{equation}
{
\begin{split}
\dfrac{dP(i,t)}{dt}&= Wq(i+ \ell-1|\underline{i-3})P(i-3,t)\\
&-Wq(i+ \ell+2|\underline{i})P(i,t). 
\end{split}
}
\label{mfeq5}
\end{equation}
In segment $II$\\
\begin{equation}
{
\begin{split}
\dfrac{dP(i,t)}{dt}&= W_sq(i+ \ell-1|\underline{i-3})P(i-3,t)\\
& -W_sq(i+ \ell+2|\underline{i})P(i,t) .
\end{split}
}
\label{mfeq5a}
\end{equation}
%%%%%%%%%%%%%%%%%%%%%%%%%%%%%%%%%%%%%%%%%%%%%

%%%%%%%%%%%%%%%%%%%%%%%%%%%%%%%%%%%%%%%%%%%%%

%%%%%%%%%%%%%%%%%%%%%%%%%%%%%%%%%%%%%%%%%%%%%%%%%%
\begin{figure}[t] 
\begin{center}
\includegraphics[width=0.85\columnwidth]{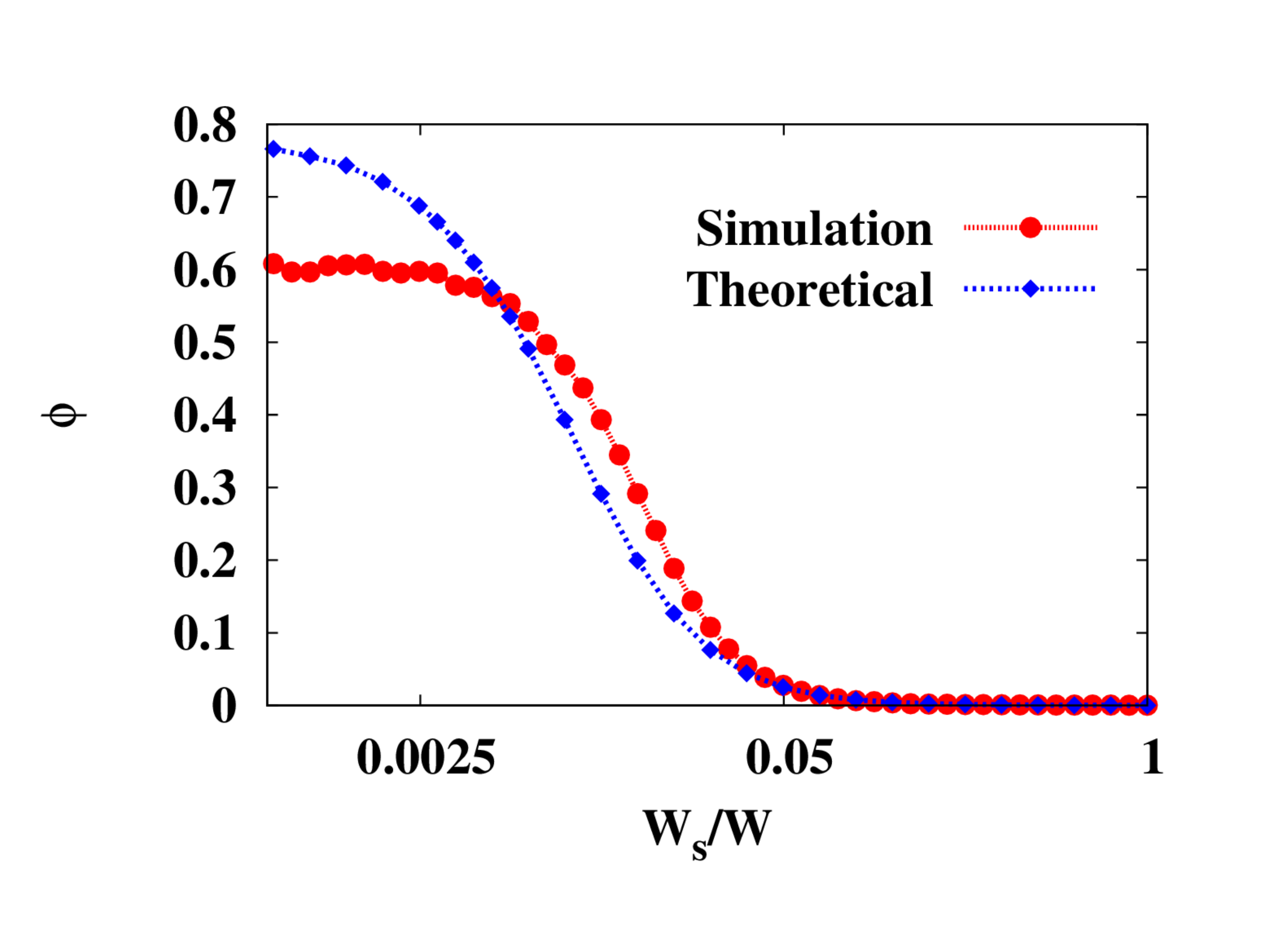}  
\end{center} 
\caption[scale=tiny]{(Color online) The fraction of proteins that are `fusion' of two proteins 
arising from PRF are plotted as a function of the jump rate ratio $ \gamma = W_{s}/W$. 
System size is $ N=1000+\ell-1 $, with the rod size $ \ell=18 $, and $W=83$ s$^{-1}$}. 
\label{fusion}
\end{figure}
%%%%%%%%%%%%%%%%%%%%%%%%%%%%%%%%%%%%%%%%%%%%%%%%%%

%%%%%%%%%%%%%%%%%%%%%%%%%%%%%%%%%%%%%%%%%%%%%

%%%%%%%%%%%%%%%%%%%%%%%%%%%%%%%%%%%%%%%%%%%%%%
\section{Results}
%%%%%%%%%%%%%%%%%%%%%%%%%%%%%%%%%%%%%%%%%%%%%%%

In this work we have followed two different approaches for analyzing the model to obtain quantitative 
results. All the effective rate constants $W$, $W_{s}$ and $W_{fs}$ chosen for calculation were 
obtained using the relation between the effective rate constant and basic rate constants of the 
5-state original model whose numerical values are given in Table.~1 of the supplementary information 
given in \cite{mishra16a}.

In our first approach we have obtained the steady state occupational density profile by solving the Eqs. (\ref{mfeq1}) - (\ref{mfeq5a}) iteratively, under mean field approximation (MFA), by the 
standard Runge-Kutta integration scheme until the steady state was attained. 
In our second approach 
we carried out Monte-Carlo simulations (MCS) of the same mechanisms as in our theoretical model. 
In both the approaches a very high initiation rate $ \alpha \approx 100s^{-1} $ and very low termination rate $ \beta\leq 10s^{-1} $ have been chosen to ensure a high density of the rods on the track.

Realistically, typical values of ${\ell}$ would be about 30 nucleotides. In order to study our model with ${\ell}=30$, we would need a proportionately large values of $L_{1}$, $L_{2}$ and $L$. A comparison of our preliminary test results obtained for 
those realistic  sizes with those for the shorter values reported in this manuscript showed no qualitative difference.Therefore, with the shorter values of the sizes, namely, $ L_{1}=399, L_{2}=450, L=1000$, we got lot 
more data for averaging which are reported in this paper.

In Fig.~(\ref{eff1}), we show the variation in frameshift flux $ J_{fs} $ with jump rate ratio $ \gamma =W_{s}/W $. 
The theoretical results predicted under MFA agree well with the corresponding data from 
MC simulation in the regime where the coverage density of the ribosomes is sufficiently low. 
However, with increasing coverage density increasing deviation of the MFA from the MC data is 
observed. This is not unexpected because the correlations that are significant at high densities 
are neglected under MFA.

From Eq.~(\ref{frameshiftfluxA}), $ J_{fs} $ depends on two factors, (a) the parameter $W_{fs}$ which, 
in turn, reflects the complexity of the pseudoknot and (b) the variable $p(L_{1}|\underline{L_{1}+1})$ 
which indicates the availability of an empty target site for frameshift. As the psudoknot strength 
$\Delta \tilde{G}$ increases the parameter $W_{fs}$ also increases (equivalently, $ \gamma $ 
decreases), thereby causing an increase of $J_{fs}$. However, the increase of  $\Delta \tilde{G}$ also 
suppresses $W_{s}$, leading to the increasingly dense queue of ribosomes just behind the special site, 
i.e., a concomitant decrease of $p(L_{1}|\underline{L_{1}+1})$. Therefore, with further increase of 
$\Delta \tilde{G}$ (i.e., decrease of $\gamma$) the flux $J_{fs}$ attains its maximum beyond which 
it decreases sharply because the target site becomes inaccessible at such high density of ribosomes.

%%%%%%%%%%%%%%%%%%%%%%%%%%%%%%%%%%%%%%%%%%%%%%%%%%

In Fig.~(\ref{eff3}), we show the occupational density profile (in steps of 3 nucleotides i.e. for $ j=1,4,7,10, \dots $) for $ W_{fs}=13.3s^{-1}$ and  $W_{s}= 0.00074s^{-1}$. For this value of $ W_{s}\ll W $ the entire segment $II$ behaves as a static bottleneck which creates a high density region in segment $I$ and a low density region in segment $III$. In the inset we show the occupational density profile (in steps of 1 nucleotide, i.e., for $ j=1,2,3,\dots $) for a small region from $ i=390 $ to $ i=420 $.  Since a rod can jump only in steps of $ 3$ nucleotides, it can occupy only those sites which are of the form $ 3k+1 $, where $ k=1,2,3,\dots $. Everywhere else occupational density will always remain zero behind the special point. After  passing through the special site, a rod can occupy site $ 3k+1 $ (non-frameshifted rod) as well as site  $ 3k $ (frameshifted rod) Thus, the structure observed in the occupational density profile is merely an artefact of the hopping of each 
rod, normally, by three sites at a time.
%%%%%%%%%%%%%%%%%%%%%%%%%%%%%%%%%%%%%%%%%%%%%%%%%%
%%%%%%%%%%%%%%%%%%%%%%%%%%%%%%%%%%%%%%%%%%%%%%%%%%

%%%%%%%%%%%%%%%%%%%%%%%%%%%%%%%%%%%%%%%%%%%%%%%%
In Fig.~(\ref{flux1}), we show the variation in flux $ J $ with coverage density $ \rho_{c} $ for three different values of $ \gamma $. The flat top of the flux is a  well-known feature of the TASEP with a blockage \cite{Jano92,Schu93a}. 

Another interesting quantity that characterizes PRF is the fraction $\phi$ of the 
proteins synthesized that are actually {\it fusion} of two proteins ``conjoined at birth'' 
because of PRF. 
Let $n_{fs}$ be the total number of ribosomes that undergo frameshift from the special 
point and detach from the site $ L-1 $ (i.e. the designated site of termination of translation 
by frameshifted ribosomes) after completing synthesis of a fusion protein without 
suffering premature detachment. Similarly, $n_{nfs}$ is the total number of ribosomes 
that complete synthesis of a full length single protein without undergoing frameshift and 
detach from the designated termination site $L$. In terms of $n_{fs}$ and $n_{nfs}$, 
the fraction $\phi$ is defined as
\begin{equation}
\phi = \dfrac{n_{fs}}{n_{fs}+n_{nfs}}
\label{phi}
\end{equation}
Thus, $\phi$ is also a measure of the efficiency of programmed -1 frameshift. 
The fraction $\phi$ is plotted against the ratio $W_{s}/W$ in Fig.\ref{fusion}. 
The deviation of the mean-field prediction from the corresponding computer 
simulation data increases with decreasing $W_{s}/W$. This deviation is caused by 
the increasing crowding of ribosomes where correlations, which are neglected in 
MFA, are quite significant. 

The fraction $\phi$ expectedly vanishes in the limit $W_{s}=W$. As $W_{s}/W$ 
decreases frameshits become more likely which is reflected in the increase of $\phi$. 
However, as the ratio $W_{s}/W$ decreases further the number of frameshift events 
$n_{fs}$ begins to decrease because of the increasing unavilability of an empty target 
site that is needed for -1 frameshift. But, concomitantly $n_{nfs}$ also decreases 
because of the inability to unwind a stiffer pseudoknot. Consequently, $\phi$ saturates  
as $\gamma = W_{s}/W \to 0$.

%%%%%%%%%%%%%%%%%%%%%%%%%%%%%%%%%%%%%%%%%%%%%%%%%%

%%%%%%%%%%%%%%%%%%%%%%%%%%%%%%%%%%%%%%%%%%%%%%%%%%
\section{Conclusion}
%%%%%%%%%%%%%%%%%%%%%%%%%%%%%%%%%%%%%%%%%%%%%%%%%%

Programmed ribosomal frameshift is one of the most prominent modes of recoding of genetic 
information. In this paper we have demonstrated that the density of ribosomes at and around 
the slippery sequence is an important parameter that determines the frequency of programmed 
ribosomal -1 frameshift. In Figs.~(\ref{eff1}) and (\ref{fusion}) we have demonstrated 
the effects by varying the stiffness of the pseudoknot that, in turn, controls the ribosome density.

The suppression of -1 programmed frameshift by a trailing ribosome in a dense ribosome traffic 
on a mRNA track is similar to the suppression of diffusive backtracking of a RNA polymerase 
(RNAP) motor by another trailing very closely on a DNA track\cite{sahoo11}. Unlike the slippery 
sequence and the pseudoknot on the mRNA template, this hitherto neglected parameter 
may control the frequency of -1 frameshift  dynamically because the density of the ribosomes 
on the mRNA can be up- or down-regulated by several different signals and pathways. 

For laboratory experiments, the stiffness of a pseudoknot can be varied artificially  
\cite{tholstrup12} .
Using such synthetic mRNA strands our theoretical prediction can be tested experimentally 
by a combination of ribosome profiling technique \cite{ingolia14,ingolia16} (for measuring the 
ribosome density) and FRET (for the frequency of frameshift) \cite{tinoco11}. 

From the perspective of TASEP-based modeling of ribosome traffic on mRNA template, our 
work is a significant progress. In all the earlier models of this type each site on the lattice 
(chain) corresponds to a single codon, i.e., a triplet of nucleotides. In contrast, in the models 
developed here each site represents a single nucleotide; this modification was necessary 
to capture -1 frameshift where the ribosome steps backward on its mRNA track by a single 
nucleotide. We intend to use this prescription for modeling the mRNA track in our future 
model of programmed +1 frameshift. The pseudoknot segment of the mRNA track is, 
effectively, an extended ``blockage'' against the forward movement of the ribosomes. 
Not surprisingly, the density profile (see Fig.\ref{eff3}) and the net flux 
(see Fig.\ref{flux1}) of the ribosomes display the well 
known characteristics of TASEP with static blockage \cite{Jano92,Schu93a}.

\section{Acknowledgments}
%%%%%%%%%%%%%%%%%%%%%%%%%%%%%%%%%%%%%%%%%%%%%
This work has been supported by J.C. Bose National Fellowship (DC), ``Prof. S. Sampath Chair'' Professorship (DC) and by UGC Junior Research Fellowship (BM).\\
\newpage
\newpage
%%%%%%%%%%%%%%%%%%%%%%%%%%%%%%%%%%%%%%%%%%%%%%
\section{Supplementary Information} 
%%%%%%%%%%%%%%%%%%%%%%%%%%%%%%%%%%%%%%%%%%%%%

\subsection{5-state model of elongation cycle} 

A ribosome consists of two interconnected subunits called large subunit (LSU)
and small subunit (SSU). A tRNA molecule can transit through the intersubunit space.
The three binding sites for the tRNA, arranged sequentially from the entry to the exit along
its path, are denoted by A, P and E, respectively. One end of the tRNA molecule that
interacts with the SSU constitutes the anticodon that binds with the codon on the mRNA
track by complementary base-pairing. The other end of the tRNA, that interacts with the
LSU, is aminoacylated, i.e., charged with the corresponding amino acid monomer.

%%%%%%%%%%%%%%%%%%%%%%%%%%%%%%%%%%%%%%%%%%%%%%%%%%%%
\begin{figure}[b!] 
\begin{center}
\includegraphics[width=0.75\columnwidth]{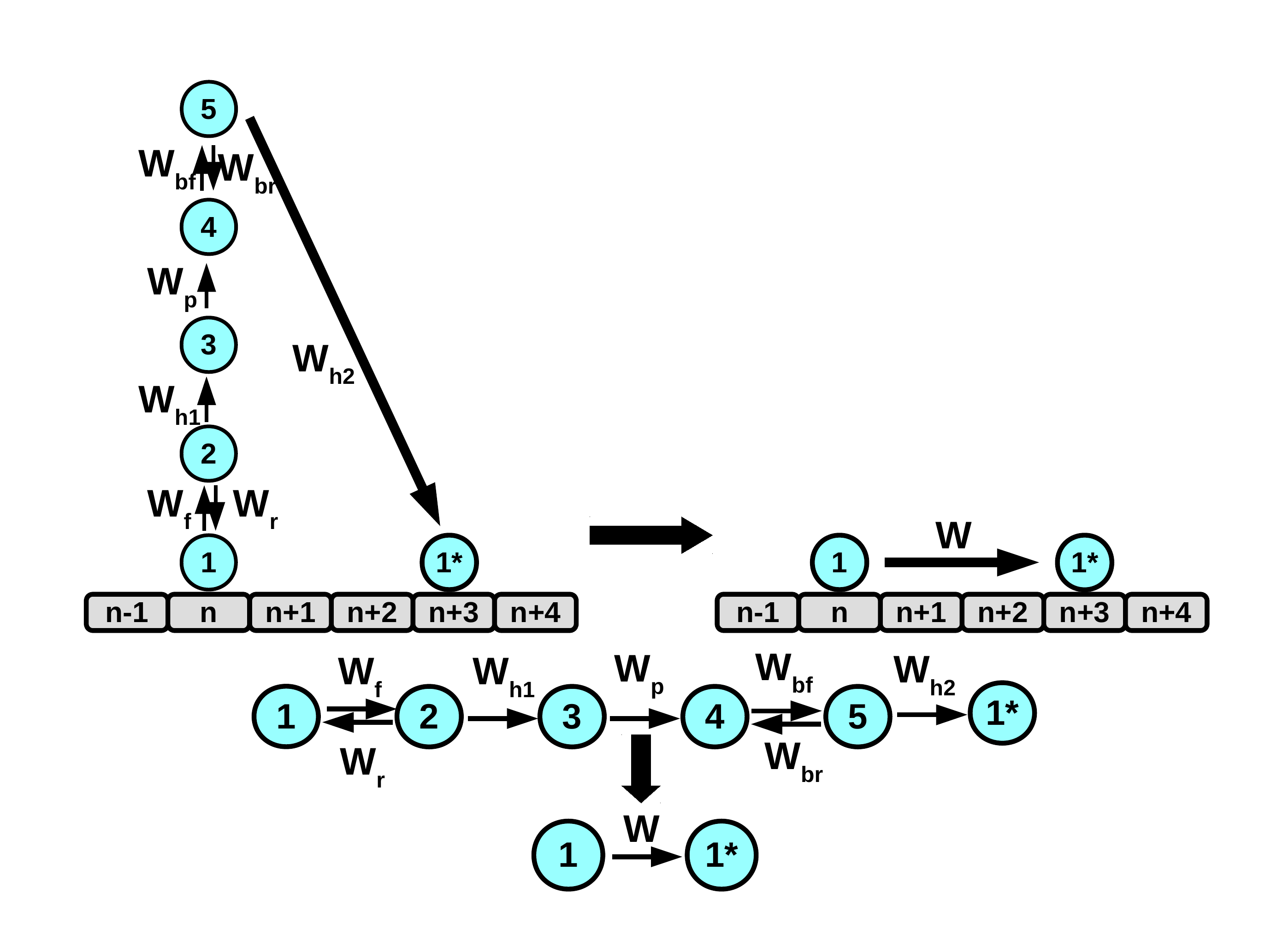}  
\end{center} 
\caption[scale=tiny]{Schematic diagram for conversion of a 5-state model into a single state model} 
\label{conversion}
\end{figure}
%%%%%%%%%%%%%%%%%%%%%%%%%%%%%%%%%%%%%%%%%%%%%%%%%%

The elongation cycle can be broadly divided into a sequence of three major processes:
aatRNA selection, peptide bond formation, and translocation. However, aatRNA selection itself proceeds in two steps: (i) Preliminary selection based on the differences in the
codon-anticodon binding free energies for cognate and non-cognate aatRNA, (ii) kinetic discrimination between cognate and near-cognate aatRNA that is usually referred to as kinetic
proofreading. Once an aatRNA passes through these two-step ``molecular identification''
screening, it is accommodated at the A-site and the system awaits the onset of the next
major process, namely, peptide bond formation. The third major process, namely translocation, also consists of multiple sub-steps during which the two ribosomal subunits execute
well orchestrated movements on the template mRNA while concomitant movement of the
tRNAs take place on the triplet binding sites E,P,A on the ribosome.

An elongation cycle is captured in our model by a multi-step kinetic process involving $N$ 
distinct mechano-chemical states. The number $N=5$ and the allowed transitions among them, 
as shown schematically in Fig. (\ref{conversion}), capture all the major steps in the elongation cycle 
that have been established so far by structural and kinetic measurements \cite{frank10}. 
The rates of the allowed transitions indicated by the arrows on  this diagram are also shown 
symbolically next to the corresponding arrows. In principle, frameshift can take place from any 
of the five states of this kinetic model. Therefore, corresponding to the 5-state model of 
elongation cycle, one can envisage at least five different kinetic models of -1 frameshift; 
the only difference between these different models is the step in which frameshift is assumed 
to take place. The details of all these five different models are given elsewhere \cite{mishra16b}. 

Although the slippery sequence and the downstream pseudoknot have been implicated in the 
-1 frameshift the exact step of the longation cycle where it occurs and the structural dynamics 
causing this frameshift remain controversial. Therefore, in this reduced minimal model the five distinct 
mechano-chemical states are collapsed onto a single effective state labelled by the instantaneous 
position of the ribosome.

%%%%%%%%%%%%%%%%%%%%%%%%%%%%%%%%%%%%%%%%%%%%%%%%%%%%%%%%%%%%%%%%%%%%%%%

%%%%%%%%%%%%%%%%%%%%%%%%%%%%%%%%%%%%%%%%%%%%%%%%%%
\subsection{Reducing the 5-state model to a 1-state model} \label{app1}
%%%%%%%%%%%%%%%%%%%%%%%%%%%%%%%%%%%%%%%%%%%%%%%%%%%%%%%%%%%%

%%%%%%%%%%%%%%%%%%%%%%%%%%%%%%%%%%%%%%%%%%%%%%%%%%%%%%%%%%%%
\begin{table}[h!]
 \begin{tabular}{ |c|c|c| }
  \hline
 \textbf{Rate Exp.} & \textbf{Num.Value} & \textbf{Source}  \\
  \hline 
  $ W_{f}$ & $ 100s^{-1} $  & Pape T. et al.\cite{pepe98}\\  
  $W_{r}$ & $25s^{-1}  $& Pape T. et al.\cite{pepe98}\\
  $W_{h1} $ & $80s^{-1}  $& Rodnina M.V. et al.\cite{rodnina95}\\
  $ W_{p} $ & $80s^{-1}  $& Pape T. et al.\cite{pepe98}\\
  $ W_{bf} $ & $25s^{-1}  $& Sharma A.K. et al.\cite{sharma11b}\\
  $W_{br}$ & $25s^{-1}  $& Sharma A.K. et al.\cite{sharma11b}\\
  $ W_{h2} $ & $60s^{-1}  $& Wen J.D. et al.\cite{wen08}\\
  \hline 
\end{tabular}
\caption{Parameters in the 5-state model of elongation cycle and their numerical values}
\label{table:A}
\end{table}
%%%%%%%%%%%%%%%%%%%%%%%%%%%%%%%%%%%%%%%%%%%%%%%%
%%%%%%%%%%%%%%%%%%%%%%%%%%%%%%%%%%%%%%%%%%%%%%%%%%
By using effective rate constant method we have converted the five state model into an equivalent one state generic model\cite{cleland75} Fig. (\ref{conversion}). The effective rate constants are,
%%%%%%%%%%%%%%%%%%%%%%%%%%%%%%%%%%%%%%%%%%

%%%%%%%%%%%%%%%%%%%%%%%%%%%%%%%%%%%%%%%%%%%
\begin{equation}
W_{4}=W_{bf}\dfrac{W_{h2}}{W_{h2}+W_{br}}
\label{rc1}
\end{equation}
\begin{equation}
W_{2}=W_{f}\dfrac{W_{h1}}{W_{h1}+W_{r}}
\label{rc2}
\end{equation}
and 
\begin{equation}
\begin{split}
[W_{1\rightarrow 1^{*}}]^{-1} &= [W]^{-1}=\dfrac{1}{W_{2}}+\dfrac{1}{W_{h1}}+\dfrac{1}{W_{p}}\\
& +\dfrac{1}{W_{4}}+\dfrac{1}{W_{h2}}.
\end{split}
\label{rc3}
\end{equation}

%%%%%%%%%%%%%%%%%%%%%%%%%%%%%%%%%%

\subsection{Calculation of the conditional probabilities $p$ and $q$} 
\label{app2}
%%%%%%%%%%%%%%%%%%%%%%%%%%%%%%%%%%

%%%%%%%%%%%%%%%%%%%%%%%%%%%%%%%%%%%%%%%%%%%%%
\begin{figure}[t] 
\begin{center}
\includegraphics[width=0.7\columnwidth]{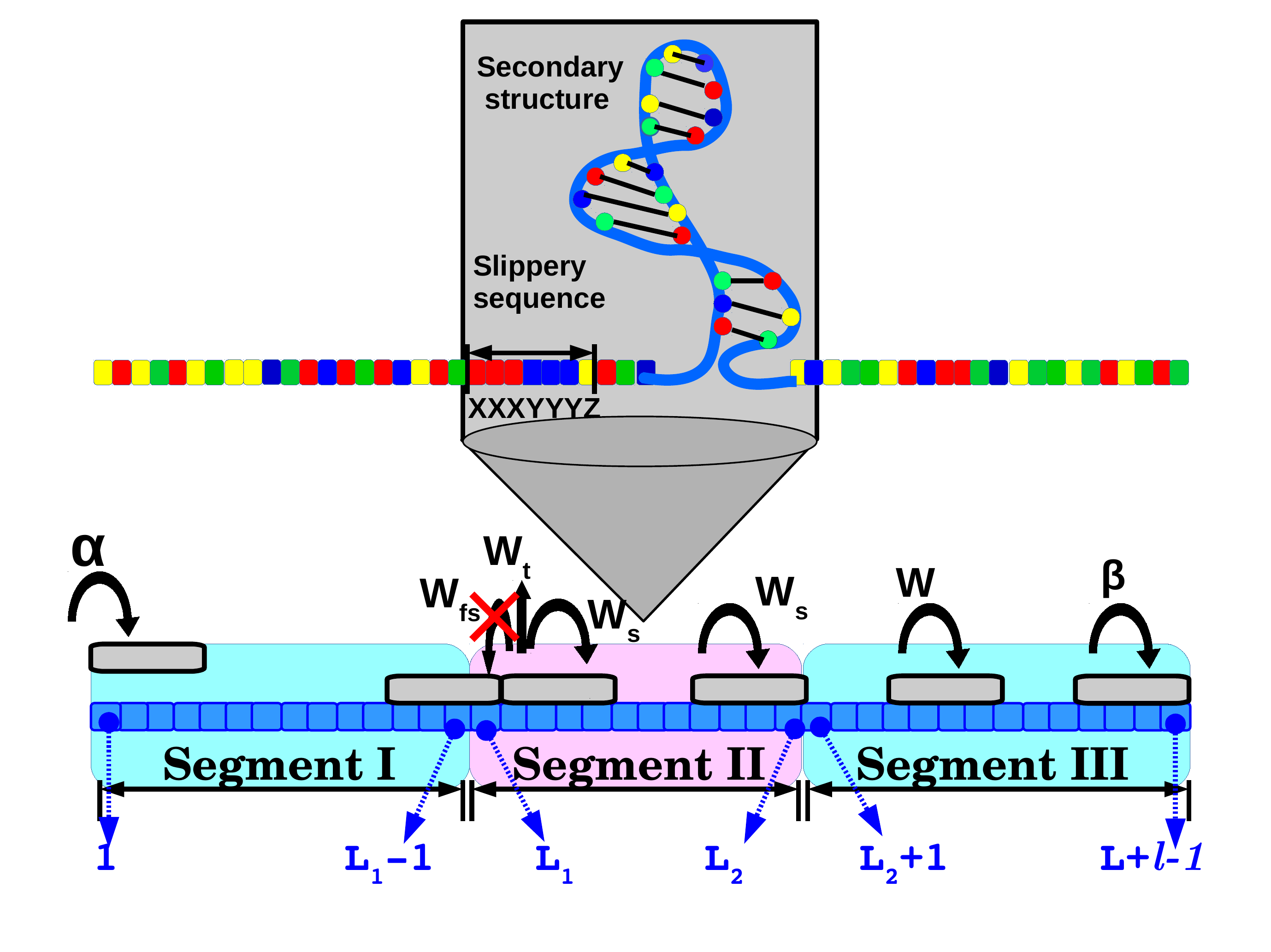}  
\end{center} 
\caption[scale=tiny]{Schematic diagram for Model B.  } 
\label{modelB}
\end{figure}
%%%%%%%%%%%%%%%%%%%%%%%%%%%%%%%%%%%%%%%%%%%

%%%%%%%%%%%%%%%%%%%%%%%%%%%%%%%%%%%%%%%%%%%%%%%%%
\begin{figure}[b!] 
\begin{center}
\includegraphics[width=0.7\columnwidth]{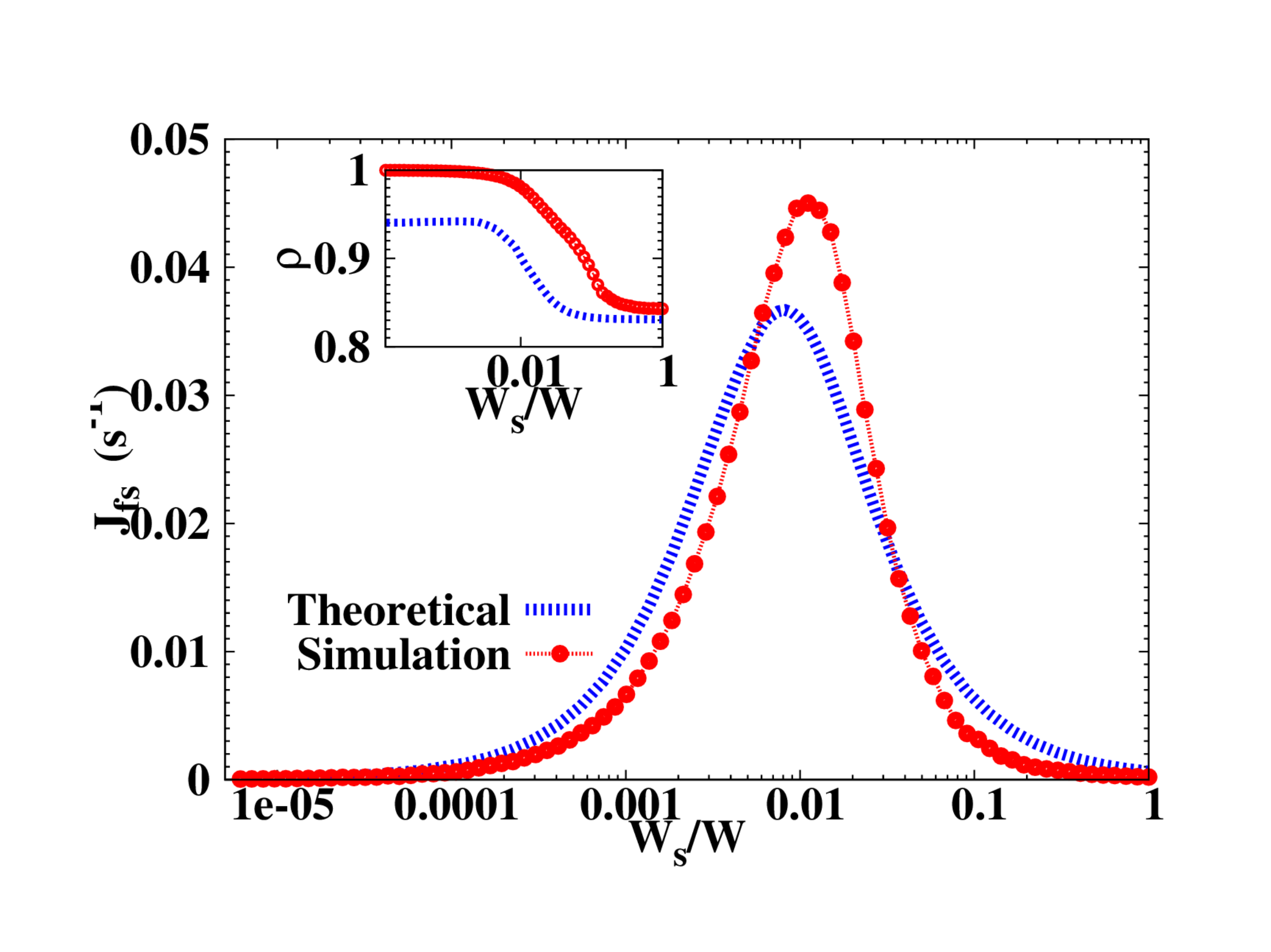}  
\end{center} 
\caption[scale=tiny]{ Frameshift flux $ J_{fs} $ for model B. In inset, coverage probability at target site $ i=L_{1} $ for frameshift. Dashed line is for theoretical values under MFA and dashed line with $\circledcirc$ is for MC simulation with initiation rate $ \alpha=100s^{-1} $, and termination rate $ \beta=10s^{-1} $. System size {$ N=L+ \ell-1=1000+\ell-1 $} and rod size {$ \ell=18 $}.} 
\label{eff4}
\end{figure}
%%%%%%%%%%%%%%%%%%%%%%%%%%%%%%%%%%%%%%%%%%%%%%%%%%

Consider $N$ rods, each of size {$\ell$}, distributed over a periodic lattice of $ L $ sites. 

We first calculate 
the conditional probability $p(i-1 \mid \underline{i})$ that the site $ i -1 $  is not covered, given that a ribosome is 
located at the site $ i $ ((i.e., covers the sites {$i,i+1,...,i+{\ell}-1$}).
In order to do so we follow \cite{shaw03,scho04} and map the process to the so-called 
zero-range process (ZRP) \cite{evans05}. In this mapping one labels the $N$ rods consecutively by integers $\alpha = 1,2,\dots , N$ and the size of the gap between two consecutive rods, i.e. the number of empty sites between them, becomes a
particle occupation number $n_\alpha$ on site $\alpha$ of the ZRP.
The total number of particles in the ZRP is thus 
{$N_{ZRP}=L-N\ell$}, corresponding to the number of vacant sites in the TASEP,
and the number of sites in the  ZRP is given by $L_{ZRP} = N$. This means that
the particle number density $\rho=N/L$ of the TASEP
is related to the particle number density $c = N_{ZRP}/L_{ZRP}$ of the ZRP by 
%$\rho=1/(c+\ell+1)$ or, conversely, 
{$c=1/\rho-\ell$}. It is convenient to work
in the grand canonical ensemble where the probability of finding
$n$ ZRP-particles on (any) given site is $P_n(z) = z^n(1-z)$ and
$z=c/(1+c)$ \cite{scho04}.

Under this mapping the conditional probability $p(i-1 \mid \underline{i})$
becomes the probability of a
finding at least on particle (gap size $>0$) on a site in the zero-range process,
i.e., $p(i-1 \mid \underline{i}) = \sum_{n=1}^\infty P_n(z) $.
Thus 
\begin{equation} 
\label{cond_p}
p(i-1 \mid \underline{i}) = z = \frac{1-\ell\rho}{1-(\ell-1)\rho}.
\end{equation}

Next we calculate the conditional probability {$q(i+\ell +2\mid \underline{i})$}
which is the conditional probability that the site {$i+ {\ell} +2$} is 
not covered, given that the site $i$ is already occupied (i.e., the 
sites {$i,i+1,i+2,...,i+{\ell}-1$} are covered). In the language of the ZRP
this is the probability of finding any occupation number larger than 2, i.e.,
{$q(i+\ell +2\mid \underline{i}) = \sum_{n=3}^\infty P_n(z)$}.
Thus
\begin{equation}
q(i+\ell +2\mid \underline{i}) = z^3 = \left(\frac{1-\ell\rho}{1-(\ell-1)\rho}\right)^3.
\label{cond_q}
\end{equation}
Models with interactions between rods that are more detailed than a hard-core repulsion
can be treated with the methods of \cite{busc00}.

%%%%%%%%%%%%%%%%%%%%%%%%%%%%%%%%%%%%%%%%%%%%%
\subsection{Model B}
%%%%%%%%%%%%%%%%%%%%%%%%%%%%%%%%%%%%%%%%%%%%%

The model B differs from model A only in the rule for detachment of a 
ribosome. More specifically, a ribosome can detach from the special 
site, with rate $W_{t}$, if and only if the target site under possible 
-1 frameshift is already covered by another ribosome 
(see Fig. (\ref{modelB})).

%%%%%%%%%%%%%%%%%%%%%%%%%%%%%%%%%%%%%%%%%%%%%%%%%%%
\begin{figure}[h] 
\begin{center}
\includegraphics[width=0.7\columnwidth]{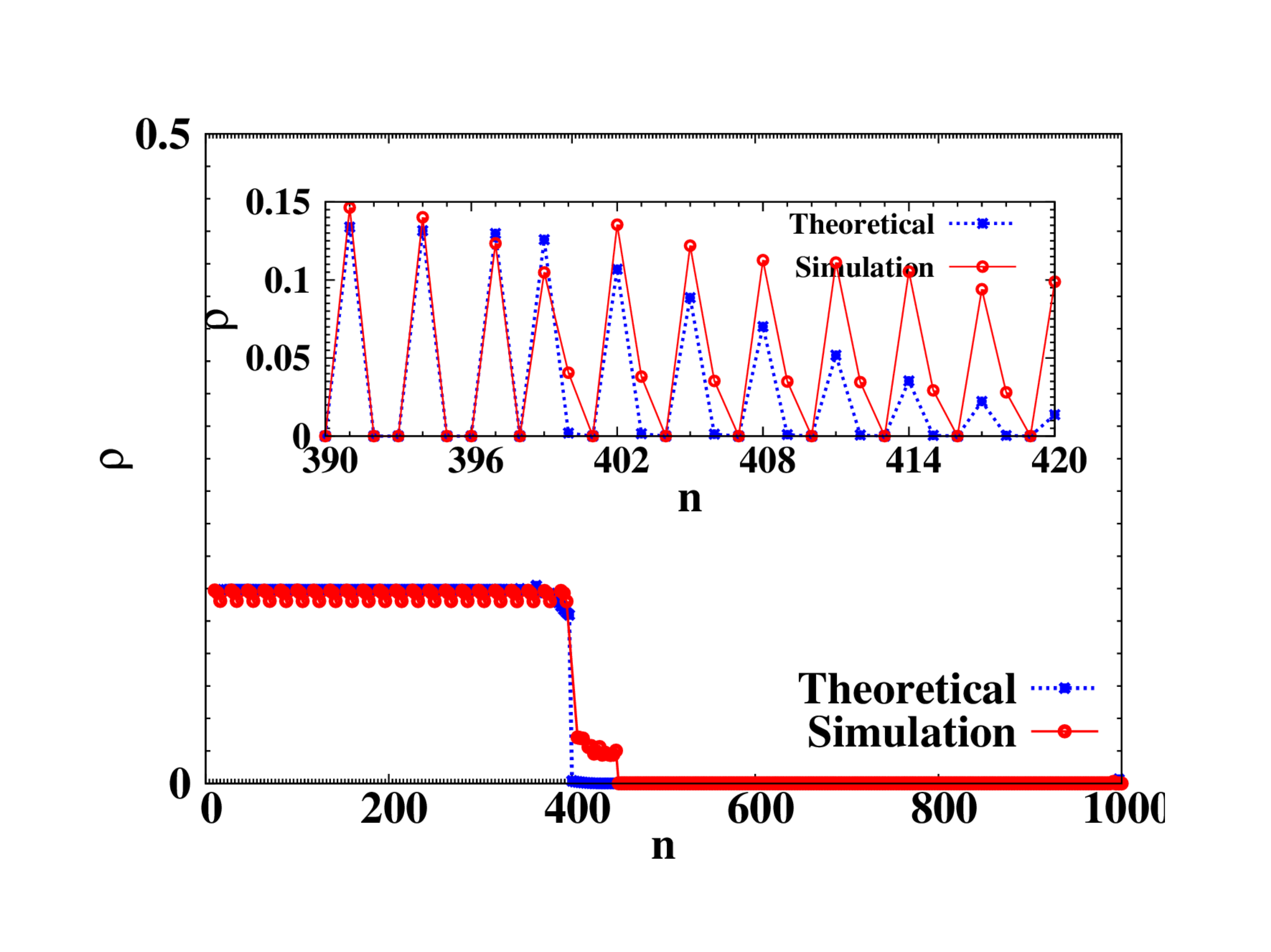}  
\end{center} 
\caption[scale=tiny]{Occupational density profile in steps of 3nts for model B. The dashed line with $\ast$ is the theoretical values under MFA and the solid line with $\odot$ is for MC simulation with initiation rate $ \alpha=100s^{-1} $, termination rate $ \beta=10s^{-1} $, $W_{fs}=13.3s^{-1}$ and $W_{s}=0.00074s^{-1}$. System size {$ L+ \ell-1=1000+\ell-1 $} and rod size {$ \ell=18 $}. The inset shows the occupational density profile of a small region from $ i=390 $ to $ i=420 $ in steps of 1 nt. The dashed line with $ \ast $ is for theoretical MF calculation and the solid line with $\odot$ is for MC simulation.} 
\label{eff5}
\end{figure}
%%%%%%%%%%%%%%%%%%%%%%%%%%%%%%%%%%%%%%%%%%%%%%%%%%

%%%%%%%%%%%%%%%%%%%%%%%%%%%%%%%%%%%%%%%%%%%%%

%%%%%%%%%%%%%%%%%%%%%%%%%%%%%%%%%%%%%%%%%%%%%%%%%

The master equations for model B are similar to those for Model A, 
except at the special site ($ i=L_{1}+1 $); at this special site 
the master equation is 
%%%%%%%%%%%%%%%%%%%%%%%%%%%%%%%%%%%%%%%%%%%%
%%%%%%%%%%%%%%%%%%%%%%%%%%%%%%%%%%%%%%%%%%%%%%%%%
\begin{equation}
\begin{split}
\dfrac{dP(i,t)}{dt} & = W_{s}q(i+ \ell-1 |\underline{i-3})P(i-3,t)  \\
& -[W_{s}q(i+ \ell+2|\underline{i})+W_{fs}p(i-1|\underline{i})]P(i,t)  \\
& -W_{t}P(i- \ell,t)P(i,t).
\end{split}
\label{mfeq_b_1}
\end{equation}
%%%%%%%%%%%%%%%%%%%%%%%%%%%%%%%%%%%%%%%%%%%%%%%%%%%%%%%%%
%%%%%%%%%%%%%%%%%%%%%%%%%%%%%%%%%%%%%%%%%%%%%%%%%%%%%%%%%

%%%%%%%%%%%%%%%%%%%%%%%%%%%%%%%%%%%%%%%%%%%%%
All the conditional probabilities for model B are identical to those 
for Model A. For this model the frameshift flux $J_{fs}$ is given by 
%%%%%%%%%%%%%%%%%%%%%%%%%%%%%%%%%%%%%%%%%%%%%%

%%%%%%%%%%%%%%%%%%%%%%%%%%%%%%%%%%%%%%%%%%%%%%%%%%%
\begin{equation}
J_{fs}=W_{fs}p(L_{1}|\underline{L_{1}+1})P(L_{1}+1,t).
\label{frameshiftfluxB}
\end{equation} 

%%%%%%%%%%%%%%%%%%%%%%%%%%%%%%%%%%%%%%%%%%%%%%%%%%
%%%%%%%%%%%%%%%%%%%%%%%%%%%%%%%%%%%%%%%%%%%%%%%%%%
\subsection{Results for model B}
%%%%%%%%%%%%%%%%%%%%%%%%%%%%%%%%%%%%%%%%%%%%%%%%%%
The results obtained for Model B are also similar to the corresponding 
results for Model A. In Fig. (\ref{eff4}) we show the variation of 
frameshift flux $ J_{fs} $ with the jump rate ratio $ \gamma $. 
In the inset we show the variation of the coverage density at the 
target site $ i=L_{1} $ with the variation of the ratio $ \gamma $.

%%%%%%%%%%%%%%%%%%%%%%%%%%%%%%%%%%%%%%%%%%%%%%%%%%%
In Fig. (\ref{eff5}) we show the coverage density profile. We observe 
a discontinuous jump between the high- and low-density phases at the 
bottleneck. Both the results are very similar to the corresponding 
results reported for model A in the main text of this letter.
%%%%%%%%%%%%%%%%%%%%%%%%%%%%%%%%%%%%%%%%%%%%%%%%%%

%%%%%%%%%%%%%%%%%%%%%%%%%%%%%%%%%%%%%%%%%%%%%%%%%%%%%%%%%%%%
%%%%%%%%%%%%%%%%%%%%%%%%%%%%%%%%%%%%%%%%%%%%%%%%%%%%%%%%%%%%%%%%%%%
%\end{appendix}

\end{document}